\documentclass{ifacconf}

\makeatletter
\let\algorithm\@undefined
\let\endalgorithm\@undefined
\makeatother

\usepackage[round]{natbib} 
\usepackage{graphicx}      
\usepackage{tikz}\usetikzlibrary{calc,patterns,
                 decorations.pathmorphing,
                 decorations.markings}
\usepackage{amsmath}
\usepackage{multirow}
\usepackage[]{algorithm2e}
\usepackage{mathtools}
\usepackage[T1]{fontenc}
\begin{document}
\begin{frontmatter}

\title{Damage Identification for The Tree-like Network through Frequency-domain Modeling\thanksref{footnoteinfo}} 

\thanks[footnoteinfo]{The partial support of NSF Award 1826079 is gratefully acknowledged.}

\author[First]{Xiangyu Ni} 
\author[Second]{Bill Goodwine} 

\address[First]{Department of Aerospace and Mechanical Engineering, University of Notre Dame, IN 46556 USA (e-mail: xni@nd.edu).}
\address[Second]{Department of Aerospace and Mechanical Engineering, University of Notre Dame, IN 46556 USA (e-mail: bill@controls.ame.nd.edu)}

\begin{abstract}                
In this paper, we propose a method to identify the damaged component and quantify its damage amount in a large network given its overall frequency response. The identification procedure takes advantage of our previous work which exactly models the frequency response of that large network when it is damaged. As a result, the test shows that our method works well when some noise present in the frequency response measurement. In addition, the effects brought by a damaged component which is located deep inside that large network are also discussed.
\end{abstract}


\end{frontmatter}

\section{Introduction}

This paper proposes a damage identification method for a tree model given a noisy measurement of its overall frequency response when there exists one damaged component. The tree model, which is shown in Fig.~\ref{fig:tree}, has many applications. For example, see its application to viscoelastic behavior in \cite{heymans1994fractal}, blood vessel in \cite{gabrys2005fractal}, the vascular tree in human retina in \cite{masters2004fractal}, and 1D relaxation of the aortic valve in \cite{doehring2005fractional}. The main reason for us to choose that model as the starting point of our work is because its transfer function is rational with real orders and commensurable (See \cite{valerio2013introduction} for the definition of commensurable transfer function), which is the simplest case among all non-integer-order systems. Note that although the tree model consists of linear springs and dampers, it can also be converted to an electrical, a fluid or a thermal system since springs and dampers have their corresponding equivalent components for those systems.

Fault detection is indispensable to modern industry in real life. Consequently, different types of fault detection methods have been proposed, for example \cite{roemer2000advanced} and \cite{sikorska2011prognostic}. One type of methods uses system identification to monitor a system's health, for example \cite{juang1985eigensystem}, \cite{brincker2001modal} and \cite{peeters2001stochastic}. The damage identification method proposed in this paper belongs to that type. In addition to using system identification, the method proposed in this paper also leverages the result from our previous work, which exactly models the frequency response of a tree with one damaged component. That knowledge from modeling brings two advantages. First, it makes our identification procedure "know" the answer when the measurement of a damaged tree's frequency response is perfect. As a result, that modeling knowledge helps our identification method work well with a very noisy measurement. Second, that modeling knowledge also helps us to cast a damage identification problem as an optimization problem where the damage case is the decision variable directly. Therefore, we can not only identify the transfer function, but, importantly, also identify the damage inside that tree directly. As we shall see later, the overall frequency responses for large networks are non-integer order naturally. Hence, our proposed method is derived from a fractional-order system identification method proposed by \cite{oustaloup1995derivation}. Other fractional-order system identification methods can be seen in \cite{hartley2003fractional}, \cite{liu2013identification}, and \cite{zhou2013genetic}.

There are at least two limitations for this work at its current state. First, the number of damaged components is limited to one in this paper. Second, it only applies to one specific network, the tree model. Therefore, we are working on generalizing this idea to multiple damaged components in a class of networks similar to the tree. The result of our initial analysis is promising.

A closely related literature, \cite{leyden2018fractional}, from our group sets a goal similar to this paper. However, they are not exactly the same. \cite{leyden2018fractional} uses the order variation in a large network's transfer function to monitor its health which is different from the method proposed in this paper. Furthermore, that work cannot identify the exact damage case inside a network. In contrast, this paper aims at locating the damaged component and quantifying its damage amount.

The rest of this paper is organized as follows. Section~\ref{sec:background} formally defines the tree model and computes its undamaged transfer function. Most importantly, it also recaps our previous work showing how to model its damaged transfer function. Then, Section~\ref{sec:id} uses the knowledge from that damage modeling to propose a damage identification algorithm. Section~\ref{sec:idResult} shows the test results for that algorithm and lists observations about the misidentified cases. Section~\ref{sec:discussion} talks about some rationales behind that identification procedure and some effects brought by a damaged component which is located at a deep generation inside the tree. Finally, Section~\ref{sec:conclusion} concludes this paper.

\section{The Tree Model}
\label{sec:background}

\begin{figure}
\centering
\begin{tikzpicture}
\tikzstyle{spring}=[thick,decorate,decoration={zigzag,pre length=0.3cm,post length=0.3cm,segment length=6}]
\tikzstyle{damper}=[thick,decoration={markings,mark connection node=dmp,mark=at position 0.5 with 
   {
    \node (dmp) [thick,inner sep=0pt,transform shape,rotate=-90,minimum
 width=15pt,minimum height=3pt,draw=none] {};
    \draw [thick] ($(dmp.north east)+(2pt,0)$) -- (dmp.south east) -- (dmp.south
 west) -- ($(dmp.north west)+(2pt,0)$);
    \draw [thick] ($(dmp.north)+(0,-5pt)$) -- ($(dmp.north)+(0,5pt)$);
   }
 },decorate]
 
\node at (0,0) (leftF) {$f$};
\filldraw[black] (0.7,0) circle (2pt) node [above] {$x_{1,1}$};
\draw[thick, -latex] (leftF.east) -- (0.7,0);
\draw[thick] (0.7,0) -- (1.1,0);
\draw[thick] (1.1,1) -- (1.1,-1);
\draw[spring] (1.1,1) -- (2.6,1) node [midway,above=1pt] {$k_{1,1}$};
\draw[damper] (1.1,-1) -- (2.6,-1) node [midway,above=6pt] {$b_{1,1}$};
\draw (2.6,1) circle (2pt) node [above] {$x_{2,1}$};
\draw (2.6,-1) circle (2pt) node [above] {$x_{2,2}$};
\draw[thick] (2.6,1) -- (3,1);
\draw[thick] (2.6,-1) -- (3,-1);
\draw[thick] (3,1.5) -- (3,0.5);
\draw[thick] (3,-0.5) -- (3,-1.5);
\draw[spring] (3,1.5) -- (4.5,1.5) node [midway,above=1pt] {$k_{2,1}$};
\draw[damper] (3,0.5) -- (4.5,0.5) node [midway,above=6pt] {$b_{2,1}$};
\draw[spring] (3,-0.5) -- (4.5,-0.5) node [midway,above=1pt] {$k_{2,2}$};
\draw[damper] (3,-1.5) -- (4.5,-1.5) node [midway,above=6pt] {$b_{2,2}$};
\draw (4.5,1.5) circle (2pt) node [above] {$x_{3,1}$};
\draw (4.5,0.5) circle (2pt) node [above] {$x_{3,2}$};
\draw (4.5,-0.5) circle (2pt) node [above] {$x_{3,3}$};
\draw (4.5,-1.5) circle (2pt) node [above] {$x_{3,4}$};
\draw[thick] (4.5,1.5) -- (5,1.5);
\draw[thick] (4.5,0.5) -- (5,0.5);
\draw[thick] (4.5,-0.5) -- (5,-0.5);
\draw[thick] (4.5,-1.5) -- (5,-1.5);
\draw[thick] (5,2) -- (5,1);
\draw[spring] (5,2) -- (6.5,2) node [midway,above=1pt] {$k_{3,1}$};
\draw[damper] (5,1) -- (6.5,1) node [midway,above=6pt] {$b_{3,1}$};
\draw (6.5,2) circle (2pt) node [above] {$x_{4,1}$};
\draw (6.5,1) circle (2pt) node [above] {$x_{4,2}$};
\node at (7,2) {$\cdots$};
\node at (7,1) {$\cdots$};
\node at (5.5,0.5) {$\cdots$};
\node at (5.5,-0.5) {$\cdots$};
\node at (5.5,-1.5) {$\cdots$};
\draw[black, very thick] (7.3,2.5) rectangle (7.5,-2) node [midway,below=2.4cm] {$x_{last}$};
\filldraw[black] (7.4,2.4) circle (2pt);
\filldraw[black] (7.4,2.2) circle (2pt);
\filldraw[black] (7.4,2) circle (2pt);
\filldraw[black] (7.4,1.8) circle (2pt);
\filldraw[black] (7.4,1.6) circle (2pt);
\filldraw[black] (7.4,1.4) circle (2pt);
\node at (7.4,1.2) {$\vdots$};
\node at (7.4,0.7) {$\vdots$};
\node at (8.2,0) (rightF) {$f$};
\draw[thick, -latex] (rightF.west) -- (7.5,0);
\end{tikzpicture}
\caption{The tree model.}
\label{fig:tree}
\end{figure}
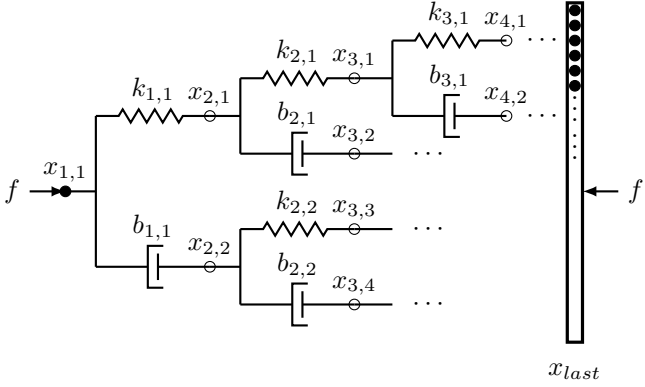

The tree model, as shown in Fig.~\ref{fig:tree}, has an infinite number of generations. At each generation, the number of nodes is doubled compared to the one on its left. For each pair of two nodes, the upper one is connected to its left node through a linear spring and the lower one is connected through a linear damper. At the last generation, all nodes are locked together. The input of interest here is the force, $f$, applied to both ends of the tree, and the output is the relative displacement, $x_{1,1}-x_\text{last}$, between both ends. Therefore, throughout this paper, without explicit exception, transfer functions always represent $(X_{1,1}(s)-X_\text{last}(s))/F(s)$.

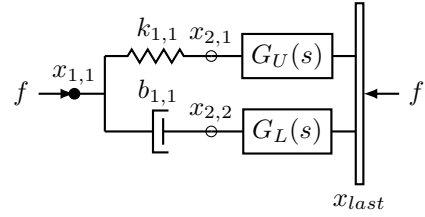
\begin{figure}
\centering
\begin{tikzpicture}
\tikzstyle{spring}=[thick,decorate,decoration={zigzag,pre length=0.3cm,post length=0.3cm,segment length=6}]
\tikzstyle{damper}=[thick,decoration={markings,mark connection node=dmp,mark=at position 0.5 with 
   {
    \node (dmp) [thick,inner sep=0pt,transform shape,rotate=-90,minimum
 width=15pt,minimum height=3pt,draw=none] {};
    \draw [thick] ($(dmp.north east)+(2pt,0)$) -- (dmp.south east) -- (dmp.south
 west) -- ($(dmp.north west)+(2pt,0)$);
    \draw [thick] ($(dmp.north)+(0,-5pt)$) -- ($(dmp.north)+(0,5pt)$);
   }
 },decorate]
 
\node at (0,0) (leftF) {$f$};
\filldraw[black] (0.7,0) circle (2pt) node [above] {$x_{1,1}$};
\draw[thick, -latex] (leftF.east) -- (0.7,0);
\draw[thick] (0.7,0) -- (1.1,0);
\draw[thick] (1.1,0.5) -- (1.1,-0.5);
\draw[spring] (1.1,0.5) -- (2.5,0.5) node [midway,above=1pt] {$k_{1,1}$};
\draw[damper] (1.1,-0.5) -- (2.5,-0.5) node [midway,above=6pt] {$b_{1,1}$};
\draw (2.5,0.5) circle (2pt) node [above] {$x_{2,1}$};
\draw (2.5,-0.5) circle (2pt) node [above] {$x_{2,2}$};
\node[draw,outer sep=0pt,thick] (M1) at (3.5,0.5) {$G_U(s)$};
\node[draw,outer sep=0pt,thick] (M2) at (3.5,-0.5) {$G_L(s)$};
\draw[thick] (2.5,0.5) -- (M1.west);
\draw[thick] (2.5,-0.5) -- (M2.west);
\draw[black, thick] (4.4,1.2) rectangle (4.5,-1.2) node [midway,below=1.2cm] {$x_{last}$};
\draw[thick] (M1.east) -- (4.4,0.5);
\draw[thick] (M2.east) -- (4.4,-0.5);
\node at (5.2,0) (rightF) {$f$};
\draw[thick, -latex] (rightF.west) -- (4.5,0);
\end{tikzpicture}
\caption{An illustration about the tree model taking advantage of its self-similarity, which is equivalent to Fig.~\ref{fig:tree}.}
\label{fig:recursive}
\end{figure}

Due to self-similarity, it is easy to see those two sub-networks, from $x_{2,1}$ and $x_{2,2}$ to $x_{\text{last}}$, are also trees. Let us use $G_U(s)$ and $G_L(s)$ to represent their transfer functions, that is
\begin{align*}
    G_U(s)&=\frac{X_{2,1}(s)-X_{\text{last}}(s)}{F_1(s)},\\
    G_L(s)&=\frac{X_{2,2}(s)-X_{\text{last}}(s)}{F_2(s)},\\
    F(s)&=F_1(s)+F_2(s).
\end{align*}
Then, the tree model can be illustratively drawn as Fig.~\ref{fig:recursive}, from which, using series and parallel connection rules for mechanical components, we can obtain that the transfer function for the entire tree can be computed by the following formula given the expressions of $G_U(s)$ and $G_L(s)$,
\begin{equation}
    G(s)=\cfrac{1}{\cfrac{1}{\cfrac{1}{k_{1,1}}+G_U(s)}+\cfrac{1}{\cfrac{1}{b_{1,1}s}+G_L(s)}},
    \label{eq:recursive}
\end{equation}
which we call the transformation formula for the tree model.

For the undamaged tree, all of its springs have the same constant $k$ and all of its dampers have the same constant $b$. In addition, we use $G_\infty(s)$ to specifically denote the transfer function for the undamaged tree model. Chapter~3 in \cite{mayes2012reduction} shows that by converting the transformation formula~(\ref{eq:recursive}) to its undamaged version, we can obtain that the undamaged transfer function for the tree model is exactly half-order:
\begin{equation}
    G_\infty(s)=\frac{1}{\sqrt{kbs}}.
    \label{eq:half}
\end{equation}

When the tree model is damaged, some of its springs' (dampers') constants are different from their undamaged value $k$ ($b$). Throughout this paper, we assume that only one component, either one spring or one damper, is damaged, which is denoted by $l$. In addition, we also assume the constant of that damaged component $l$ changes from the undamaged value $k$ ($b$) to $k\cdot\epsilon$ ($b\cdot\epsilon$), where $\epsilon$ is called its damage amount and $0<\epsilon<1$. Note that, in the rest of this paper, we use a pair ($l$,$\epsilon$) to refer to a certain damage case.

Our previous work shows that the transfer function of a damage case ($l$,$\epsilon$) for the tree model with one damaged component can be expressed as
\begin{equation*}
    G_{(l,\epsilon)}(s)=G_\infty(s)\cdot\Delta_{(l,\epsilon)}(s),
\end{equation*}
where
\begin{equation}
    \Delta_{(l,\epsilon)}(s)=\frac{N_{(l,\epsilon)}(s)}{D_{(l,\epsilon)}(s)}=\frac{\prod_{j=1}^{2g}\left(s^\frac{1}{2}+z_{j}(l,\epsilon)\right)}{\prod_{j=1}^{2g}\left(s^\frac{1}{2}+p_{j}(l,\epsilon)\right)}.
    \label{eq:delta}
\end{equation}
Here, $-z_{j}$ and $-p_{j}$ are called half-order zeros and poles. Moreover, $g$ denotes the generation where the damaged component $l$ is located. For example, when the damper $b_{3,1}$ is damaged, $l=b_{3,1}$ and thus $g=3$. Therefore, all damage cases ($b_{3,1}$, $\epsilon$) have $6$ pairs of half-order zeros and poles. Note that the damaged transfer function~(\ref{eq:delta}) is rational and commensurable.

To compute $z_j$ and $p_j$ for a specific damage case ($l$,$\epsilon$) numerically, we have to start at the first generation and go through each generation until the damaged component $l$ is reached. Such computation takes advantage of the fact that the tree model is self-similar.

As a concrete example, let us show how to use the transformation formula~(\ref{eq:recursive}) repeatedly to obtain $z_j$ and $p_j$ for the damage case ($k_{3,1}$,$\epsilon$). Due to self-similarity, from Fig.~\ref{fig:tree}, we observe that the $k_{3,1}$ component of the entire tree model is equivalent to the $k_{2,1}$ component within the sub-network from $x_{2,1}$ to $x_\text{last}$, and the $k_{2,1}$ component of the entire tree model is equivalent to the $k_{1,1}$ component within the sub-network from $x_{2,1}$ to $x_\text{last}$. Therefore, using such self-similarity, we see that we need to first compute the transfer function for the damage case ($k_{1,1}$,$\epsilon$), namely $G_{(k_{1,1},\epsilon)}(s)$. Then, using that result, we can compute $G_{(k_{2,1},\epsilon)}(s)$, which is then used to finally compute $G_{(k_{3,1},\epsilon)}(s)$.

For the damage case ($k_{1,1}$, $\epsilon$), that is when $k_{1,1}$'s spring constant becomes $k\cdot\epsilon$ and all the other springs' (dampers') constants stay at $k$ ($b$), the transfer function for the sub-network from $x_{2,1}$ to $x_\text{last}$ in this damage case is same as the undamaged transfer function, \textit{i.e.} $G_U(s)=G_\infty(s)$. For the same reason, $G_L(s)=G_\infty(s)$, too. Hence, for this damage case, we can replace ($G(s)$, $G_U(s)$, $G_L(s)$, $k_{1,1}$, $b_{1,1}$) with ($G_{(k_{1,1},\epsilon)}(s)$, $G_\infty(s)$, $G_\infty(s)$, $k\cdot\epsilon$, $b$) in Eq.~(\ref{eq:recursive}) which leads to
\begin{equation*}
    G_{(k_{1,1},\epsilon)}(s)=\cfrac{1}{\cfrac{1}{\cfrac{1}{k\cdot\epsilon}+G_\infty(s)}+\cfrac{1}{\cfrac{1}{bs}+G_\infty(s)}}.
\end{equation*}
After simplification, this gives
\begin{align}
    &G_{(k_{1,1},\epsilon)}(s)=G_\infty(s)\cdot\Delta_{(k_{1,1},\epsilon)}(s)\nonumber\\
    &=G_\infty(s)\cdot\frac{N_{(k_{1,1},\epsilon)}(s)}{D_{(k_{1,1},\epsilon)}(s)}\nonumber\\
    &=G_\infty(s)\nonumber\\
    &\cdot\frac{\left(s^\frac{1}{2}+\sqrt{\frac{k}{b}}\right)\left(s^\frac{1}{2}+\epsilon\sqrt{\frac{k}{b}}\right)}{\left(s^\frac{1}{2}+\epsilon\sqrt{\frac{k}{b}}+\sqrt{\frac{\epsilon(\epsilon-1)k}{b}}\right)\left(s^\frac{1}{2}+\epsilon\sqrt{\frac{k}{b}}-\sqrt{\frac{\epsilon(\epsilon-1)k}{b}}\right)}.
    \label{eq:k11}
\end{align}

Using the result in Eq.~(\ref{eq:k11}), we can now compute the transfer function for the damage case ($k_{2,1}$,$\epsilon$), that is $G_{(k_{2,1},\epsilon)}(s)$. For the similar reason explained above, $G_{(k_{2,1},\epsilon)}(s)$ can be obtained by replacing ($G(s)$, $G_U(s)$, $G_L(s)$, $k_{1,1}$, $b_{1,1}$) with ($G_{(k_{2,1},\epsilon)}(s)$, $G_{(k_{1,1},\epsilon)}(s)$, $G_\infty(s)$, $k$, $b$) in Eq.~(\ref{eq:recursive}), which gives
\begin{equation*}
    G_{(k_{2,1},\epsilon)}(s)=\cfrac{1}{\cfrac{1}{\cfrac{1}{k}+G_{(k_{1,1},\epsilon)}(s)}+\cfrac{1}{\cfrac{1}{bs}+G_\infty(s)}}.
\end{equation*}
After simplification, this gives
\begin{equation}
    G_{(k_{2,1},\epsilon)}(s)=G_\infty(s)\cdot\frac{sD+s^\frac{1}{2}\sqrt{\frac{k}{b}}(N+D)+\frac{k}{b}N}{sD+s^\frac{1}{2}\sqrt{\frac{k}{b}}(N+D)+\frac{k}{b}D}.
    \label{eq:k21_1}
\end{equation}
Note that here $D=D_{(k_{1,1},\epsilon)}(s)$ and $N=N_{(k_{1,1},\epsilon)}(s)$ are known from Eq.~(\ref{eq:k11}). Hence, by using a numerical equation solver, we can find the values for $s^\frac{1}{2}$ which make either the numerator or the denominator in Eq.~(\ref{eq:k21_1}) equal to $0$, and those values are the half-order zeros and poles for damage case ($k_{2,1}$,$\epsilon$). Then, from the numerical value of those half-order zeros and poles, we can construct that
\begin{equation}
    G_{(k_{2,1},\epsilon)}(s)=G_\infty(s)\cdot\frac{\prod_{j=1}^{4}\left(s^\frac{1}{2}+z_{j}(k_{2,1},\epsilon)\right)}{\prod_{j=1}^{4}\left(s^\frac{1}{2}+p_{j}(k_{2,1},\epsilon)\right)}.
    \label{eq:k21_2}
\end{equation}

Finally, we can compute half-order zeros and poles for the damage case ($k_{3,1}$, $\epsilon$) by using that $G_{(k_{2,1},\epsilon)}(s)$ in Eq.~(\ref{eq:k21_2}). For the similar reason explained above, we can obtain $G_{(k_{3,1},\epsilon)}(s)$ by replacing ($G(s)$, $G_U(s)$, $G_L(s)$, $k_{1,1}$, $b_{1,1}$) with ($G_{(k_{3,1},\epsilon)}(s)$, $G_{(k_{2,1},\epsilon)}(s)$, $G_\infty(s)$, $k$, $b$) in Eq.~(\ref{eq:recursive}), which gives an expression similar to Eq.~(\ref{eq:k21_1}). Then, again, using a numerical equation solver, we are able to find those half-order zeros and poles for $G_{(k_{3,1},\epsilon)}(s)$.

In summary, to compute $z_j$ and $p_j$ for a damage case ($l$,$\epsilon$), we need to compute its corresponding damage case at each generation from the first one to the one where the damaged component $l$ is located. Table~\ref{tab:4Gen} shows such correspondence among all springs up to the fourth generation. For example, if we want to compute $z_j$ and $p_j$ for the damage case ($k_{4,7}$, $\epsilon$), using Table~\ref{tab:4Gen}, we know that we first need to compute $G_{(k_{1,1},\epsilon)}(s)$. Next, we need to use that result to compute $G_{(k_{2,1},\epsilon)}(s)$, which enables us to further obtain $G_{(k_{3,3},\epsilon)}(s)$. Then, we finally reach at $G_{(k_{4,7},\epsilon)}(s)$. Note that the correspondence among all dampers works exactly the same.

Additionally, as described in the example above, moving between corresponding components at two consecutive generations requires to substitute correct elements into ($G(s)$, $G_U(s)$, $G_L(s)$, $k_{1,1}$, $b_{1,1}$) in Eq.~(\ref{eq:recursive}). Such substitution is listed in Table~\ref{tab:1MoreGen}. For example, if we want to compute $G_{(k_{3,3},\epsilon)}(s)$ based on $G_{(k_{2,1},\epsilon)}(s)$, from Table~\ref{tab:1MoreGen}, we know that we can replace ($G(s)$, $G_U(s)$, $G_L(s)$, $k_{1,1}$, $b_{1,1}$) with ($G_{(k_{3,3},\epsilon)}(s)$, $G_\infty(s)$, $G_{(k_{2,1},\epsilon)}(s)$, $k$, $b$) in Eq.~(\ref{eq:recursive}). Again, note that the substitution for all dampers are exactly the same except for $b_{1,1}$ which has also been listed in Table~\ref{tab:1MoreGen}.

\begin{table}
    \centering
    \caption{Correspondence among all springs up to the $4^{th}$ generation.}
    \begin{tabular}{|c|c|c|c|}
        \hline
        $1^\text{st}$ Gen. & $2^\text{nd}$ Gen. & $3^\text{rd}$ Gen. & $4^\text{th}$ Gen. \\ \hline
        \multirow{8}{*}{$k_{1,1}$} & \multirow{4}{*}{$k_{2,1}$} & \multirow{2}{*}{$k_{3,1}$} & {$k_{4,1}$} \\ \cline{4-4}
         &  &  & {$k_{4,5}$} \\ \cline{3-4}
         &  & \multirow{2}{*}{$k_{3,3}$} & {$k_{4,3}$} \\ \cline{4-4}
         &  &  & {$k_{4,7}$} \\ \cline{2-4}
         & \multirow{4}{*}{$k_{2,2}$} & \multirow{2}{*}{$k_{3,2}$} & {$k_{4,2}$} \\ \cline{4-4}
         &  &  & {$k_{4,6}$} \\ \cline{3-4}
         &  & \multirow{2}{*}{$k_{3,4}$} & {$k_{4,4}$} \\ \cline{4-4}
         &  &  & {$k_{4,8}$} \\ \hline
    \end{tabular}
    \label{tab:4Gen}
\end{table}

\begin{table}
    \centering
    \caption{Elements substituted into ($G(s)$, $G_U(s)$, $G_L(s)$, $k_{1,1}$, $b_{1,1}$) in Eq.~(\ref{eq:recursive}) enable us to move between corresponding components at two consecutive generations.}
    \begin{tabular}{|c|c|}
        \hline
        \multirow{2}{*}{Elements substituted into Eq.~(\ref{eq:recursive})} & Corresponding \\
         & components \\ \hline
        ($G_\infty(s)$, $G_\infty(s)$, $G_\infty(s)$, $k$, $b$) & Undamaged \\ \hline
        ($G_{(k_{1,1},\epsilon)}(s)$, $G_\infty(s)$, $G_\infty(s)$, $k\cdot\epsilon$, $b$) & Undamaged$\rightarrow k_{1,1}$ \\ \hline
        ($G_{(b_{1,1},\epsilon)}(s)$, $G_\infty(s)$, $G_\infty(s)$, $k$, $b\cdot\epsilon$) & Undamaged$\rightarrow b_{1,1}$ \\ \hline
        ($G_{(k_{2,1},\epsilon)}(s)$, $G_{(k_{1,1},\epsilon)}(s)$, $G_\infty(s)$, $k$, $b$) & $k_{1,1}\rightarrow k_{2,1}$ \\ \hline
        ($G_{(k_{2,2},\epsilon)}(s)$, $G_\infty(s)$, $G_{(k_{1,1},\epsilon)}(s)$, $k$, $b$) & $k_{1,1}\rightarrow k_{2,2}$ \\ \hline
        ($G_{(k_{3,1},\epsilon)}(s)$, $G_{(k_{2,1},\epsilon)}(s)$, $G_\infty(s)$, $k$, $b$) & $k_{2,1}\rightarrow k_{3,1}$ \\ \hline
        ($G_{(k_{3,2},\epsilon)}(s)$, $G_{(k_{2,2},\epsilon)}(s)$, $G_\infty(s)$, $k$, $b$) & $k_{2,2}\rightarrow k_{3,2}$ \\ \hline
        ($G_{(k_{3,3},\epsilon)}(s)$, $G_\infty(s)$, $G_{(k_{2,1},\epsilon)}(s)$, $k$, $b$) & $k_{2,1}\rightarrow k_{3,3}$ \\ \hline
        ($G_{(k_{3,4},\epsilon)}(s)$, $G_\infty(s)$, $G_{(k_{2,2},\epsilon)}(s)$, $k$, $b$) & $k_{2,2}\rightarrow k_{3,4}$ \\ \hline
    \end{tabular}
    \label{tab:1MoreGen}
\end{table}

\section{Damage Identification Algorithm}
\label{sec:id}

In this section, we formally describe our damage identification procedure. The goal is that, for a tree with one damaged component, given a noisy measurement of its frequency response, we want to identify the damaged component and quantify its damage amount. The knowledge brought by the exact modeling of damaged trees allows us to formulate a damage identification problem as an optimization problem where the damage case $(l,\epsilon)$ is the decision variable directly. That is, we are able to come up with a metric which directly maps a damage case $(l,\epsilon)$ to a quantification of the difference between its computed frequency response and the measured one. Inspired by Chapter~4 in \cite{leyden2018system}, we use the following metric to quantify that identification error
\begin{equation}
    J(l,\epsilon)=\sum_{s}\frac{\|\Delta_{(l,\epsilon)}(s)-\overline{\Delta}(s)\|}{\|\overline{\Delta}(s)\|},
    \label{eq:idError}
\end{equation}
where $\Delta_{(l,\epsilon)}(s)$ is the computed frequency response for a damage case $(l,\epsilon)$ and $\overline{\Delta}(s)$ is a noisy measurement waiting for identification. In addition, the summation is over $s=i\cdot\omega$ where $\omega$ are the angular frequencies at which $\overline{\Delta}(s)$ is sampled.

Here is the main idea of our identification procedure. Before identification, we need a prior knowledge about where that damage is possibly located. For example, we assume that the damaged component is within the first four generations. Then, we can define a finite set for possible damaged components, that is
\begin{equation}
    L=[k_{1,1},b_{1,1},k_{2,1},b_{2,1},k_{2,2},b_{2,2},\dots,k_{4,8},b_{4,8}].
    \label{eq:ordered}
\end{equation}
For each component $l\in L$, we find the locally best $\epsilon$ which minimizes the identification error $J(l,\epsilon)$. Then, among all those locally best pairs of ($l$,$\epsilon$), we pick the one which gives the globally smallest identification error to be the final identification result ($l^*$,$\epsilon^*$).

At every $l\in L$, we solve the nonlinear programming problem
\begin{equation}
    \min_\epsilon J(l,\epsilon)=\sum_s\frac{\|\Delta_{(l,\epsilon)}(s)-\overline{\Delta}(s)\|}{\|\overline{\Delta}(s)\|},
    \label{eq:optimization}
\end{equation}
subject to
\begin{align*}
    &0<\epsilon<1; \\
    &\Delta_{(l,\epsilon)}(s)=\frac{\prod_{j=1}^{2g}\left(s^\frac{1}{2}+z_{j}(l,\epsilon)\right)}{\prod_{j=1}^{2g}\left(s^\frac{1}{2}+p_{j}(l,\epsilon)\right)}.
\end{align*}
Algorithm~\ref{alg:1} shows the pseudocode for our identification procedure. Note that $J_{\min}$ is initialized to $+\infty$ which means the largest real number determined by the machine in use.

\begin{algorithm}
    \SetAlgoLined
    \KwResult{Identify the damaged component $l^*$ and quantify its damage amount $\epsilon^*$ given the measured $\overline{\Delta}(s).$}
    $J_{\min}\leftarrow+\infty$\;
    \For{$l\in L$}
        {\For{The initial guess $\epsilon_0\in[0.1,~0.2,~\dots,~0.9]$}
            {Find ($\epsilon$,$J$) such that $\epsilon$ solves the optimization problem~(\ref{eq:optimization}) at the component $l$, and $J$ is the corresponding optimized identification error\;
            \If{$J<J_{\min}$}
                {$J_{\min}\leftarrow J$\;
                $l^*\leftarrow l$\;
                $\epsilon^*\leftarrow\epsilon$\;}}}
    \caption{Pseudocode for our identification procedure. Note that the double \texttt{for} loop here can be easily implemented in parallel.}
    \label{alg:1}
\end{algorithm}

\begin{figure}
    \centering
    \includegraphics[width=0.48\textwidth]{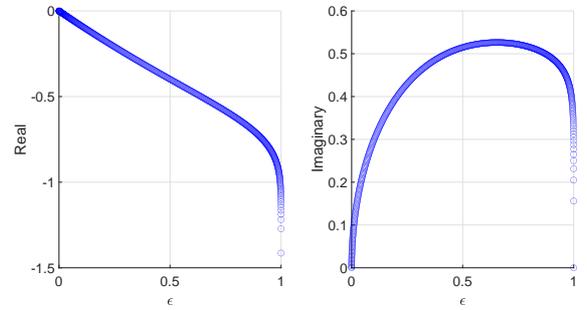}
    \caption{Sampling of the real and imaginary part for the half-order zero $-z_{7}(k_{4,7},\epsilon_a)$ where $a=1,2,\dots,500$ and $\epsilon_a$ is determined by Eq.~(\ref{eq:cheb}).}
    \label{fig:k42_z7}
\end{figure}

\begin{figure}
    \centering
    \includegraphics[width=0.48\textwidth]{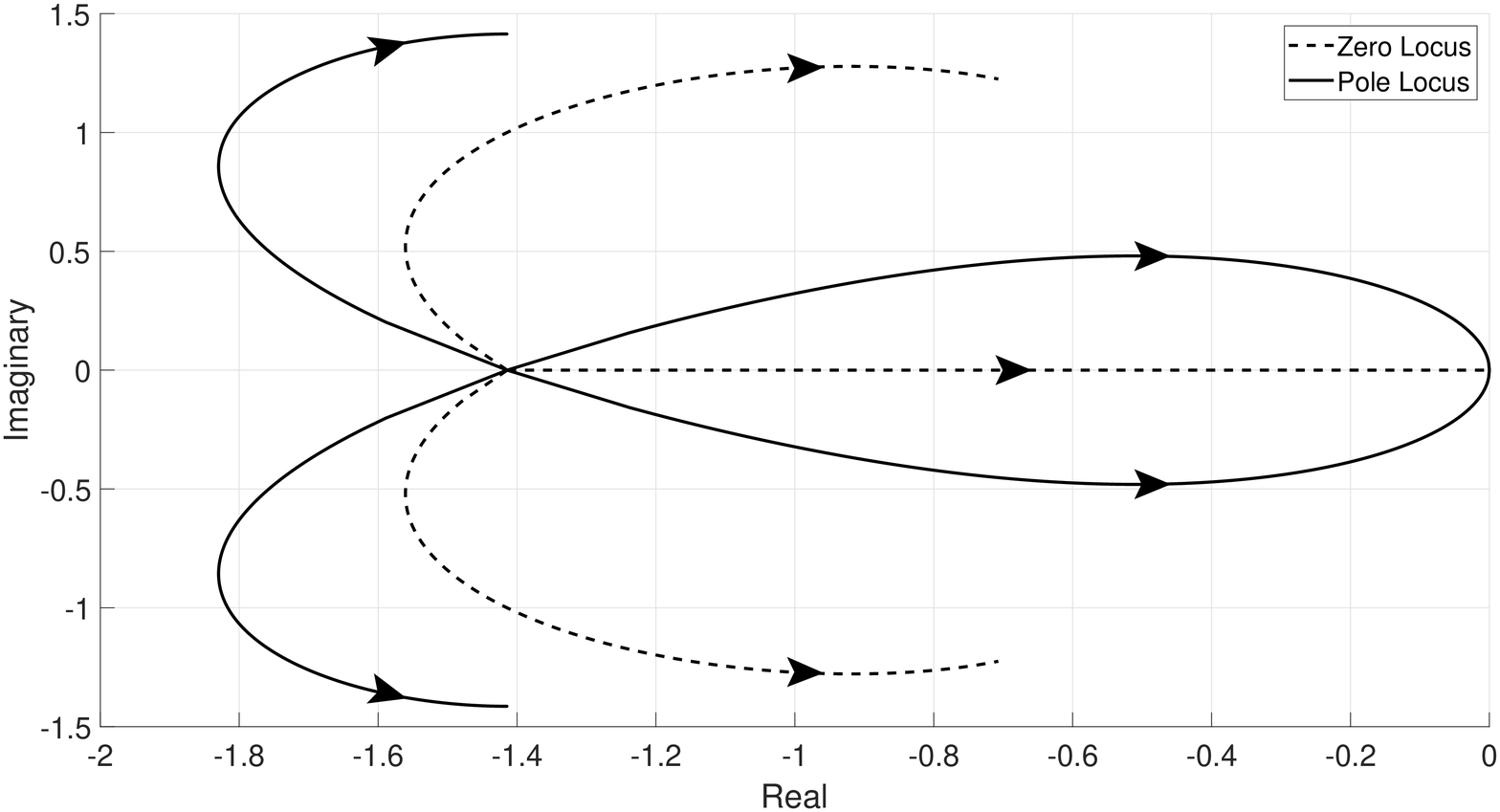}
    \caption{Locus for half-order zeros and poles, $-z_{j}(k_{2,1},\epsilon)$ and $-p_{j}(k_{2,1},\epsilon)$ for $j=1,2,3,4$ and $\epsilon$ varying from 1 (undamaged) to 0 (completely damaged). $-z_{1}(k_{2,1},\epsilon)$ stays at $-\sqrt{\frac{k}{b}}$ for all $0<\epsilon<1$.}
    \label{fig:locus_k21}
\end{figure}

Those half-order zeros and poles $-z_j$ and $-p_j$ are computed offline. Before the identification procedure shown in Algorithm~\ref{alg:1}, we sample $-z_{j}$ and $-p_{j}$ at different $\epsilon_a$ and store them to a database. Then, during the identification, we use the piecewise linear interpolation to obtain $-z_{j}$ and $-p_{j}$ from those stored values for any $0<\epsilon<1$. Specifically, at each component $l\in L$, we pick $500$ $\epsilon_a$'s between $0$ and $1$. Then, for each pair of ($l$, $\epsilon_a$), we use the method described in Section~\ref{sec:background} to compute the corresponding $-z_{j}(l,\epsilon_a)$ and $-p_{j}(l,\epsilon_a)$, and store them to a database. For example, Fig.~\ref{fig:k42_z7} shows the real and imaginary part of the half-order zero $-z_{7}(k_{4,7},\epsilon_a)$ for $a=1,2,\dots,500$.
Note that, for the damaged tree model, $-z_j$ and $-p_j$ are usually very sensitive as $\epsilon\rightarrow0$ and $\epsilon\rightarrow1$. Therefore, we pick discrete $\epsilon_a$ based on the Chebyshev nodes, that is
\begin{equation}
    \epsilon_a=\frac{1}{2}\left[\cos\left(\frac{2a-1}{1000}\pi\right)+1\right],~~a=1,2,\dots,500.
    \label{eq:cheb}
\end{equation}
Then, during the identification, we use piecewise linear interpolation to obtain half-order zeros and poles, $-z_{j}$ and $-p_{j}$, for any $0<\epsilon<1$. That is,
\begin{align}
    -z_{j}(l,\epsilon)=&-z_{j}(l,\epsilon_a)\cdot\left(1-\frac{\epsilon-\epsilon_a}{\epsilon_{a+1}-\epsilon_a}\right)\nonumber\\
    &-z_{j}(l,\epsilon_{a+1})\cdot\frac{\epsilon-\epsilon_a}{\epsilon_{a+1}-\epsilon_a},\label{eq:zContinuous}\\
    -p_{j}(l,\epsilon)=&-p_{j}(l,\epsilon_a)\cdot\left(1-\frac{\epsilon-\epsilon_a}{\epsilon_{a+1}-\epsilon_a}\right)\nonumber\\
    &-p_{j}(l,\epsilon_{a+1})\cdot\frac{\epsilon-\epsilon_a}{\epsilon_{a+1}-\epsilon_a},\label{eq:pContinuous}
\end{align}
where $\epsilon_a\leq\epsilon<\epsilon_{a+1}$. For example, Fig.~\ref{fig:locus_k21} shows the locus of half-order zeros and poles when the damaged component $l=k_{2,1}$ after interpolation. Those arrows in Fig.~\ref{fig:locus_k21} indicate the direction along which $-z_{j}(k_{2,1},\epsilon)$ and $-p_{j}(k_{2,1},\epsilon)$ move when $\epsilon$ varies from $1$ (undamaged) to $0$ (completely damaged).

\section{Identification Test Results}
\label{sec:idResult}

We test our identification procedure on all damage cases where the damaged component is located in the first three generations and each damaged component has ten different damage amounts. That is, each damage case ($l$,$\epsilon$) during the test is an element of the following Cartesian product
\begin{equation*}
    \{k_{1,1},~b_{1,1},~\dots,~k_{3,4},~b_{3,4}\}\times\{0.05,~0.15,~\dots,~0.95\}.
\end{equation*}
Therefore, 140 different damage cases are tested in total. During the test, we use \texttt{fmincon()} from \texttt{MATLAB} to solve the nonlinear programming problem~(\ref{eq:optimization}).

\begin{figure}
    \centering
    \includegraphics[width=0.48\textwidth]{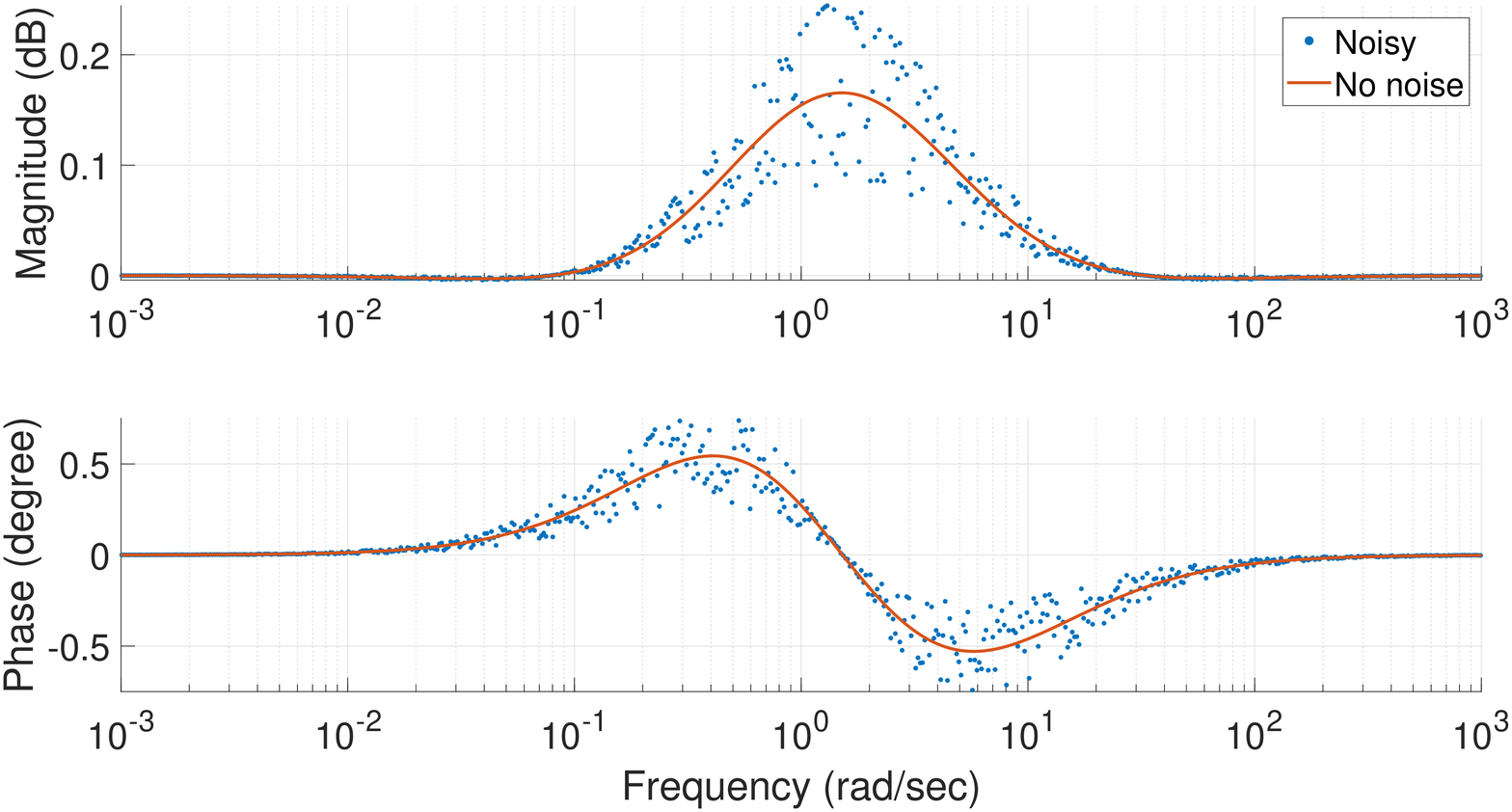}
    \caption{Bode plot of $\overline{\Delta}(s)$ to which $50\%$ noise is added when the damage case $(l,\epsilon)=(k_{3,2},0.5)$.}
    \label{fig:bode_k32_50_n50}
\end{figure}

To imitate real measurements, we add noise to the analytical value of $\overline{\Delta}(s)$. Here is what we mean by adding $n_{\max}\%$ noise to $\overline{\Delta}(s)$: If the analytical value of $\overline{\Delta}(s)=A+i\cdot B$ at some angular frequency $\omega$, what the identification procedure can see is its corresponding noisy value of $\overline{\Delta}(s)$ where
\begin{align*}
    |\overline{\Delta}(s)|&=10^{(1+n\%)\cdot\log_{10}(\sqrt{A^2+B^2})},\\
    \angle\overline{\Delta}(s)&=(1+n\%)\cdot\text{atan2}(B,A),
\end{align*}
and $n$ is a uniformly distributed random variable between $-n_{\max}$ and $n_{\max}$. For example, Fig.~\ref{fig:bode_k32_50_n50} shows the Bode plot of $\overline{\Delta}(s)$ to which $50\%$ noise is added when the damage case $(l,\epsilon)=(k_{3,2},0.5)$. Note that, by doing so, the identification result depends on the value of the random variable $n$ chosen by \texttt{MATLAB}. As a result, to accommodate that dependence, we test our identification algorithm ten times at each level of added noise.

When no noise presents in the measured $\overline{\Delta}(s)$, that is when the measured $\overline{\Delta}(s)$ is same as its theoretical value, our method correctly identifies all 140 damage cases. The maximum absolute difference between the actual damage amount $\epsilon$ and the identified $\epsilon^*$ is $1.89\times10^{-5}$ which happens at the damage case $(l,\epsilon)=(b_{3,1},0.45)$.

\begin{figure}
    \centering
    \includegraphics[width=0.48\textwidth]{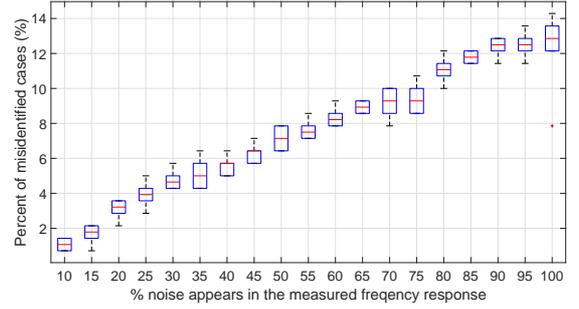}
    \caption{Box plot for the percentage of misidentified cases out of total 140 damage cases \textit{versus} the percentage of noise added to $\overline{\Delta}(s)$. Ten runs for each level of added noise.}
    \label{fig:numMidIDVsNoise}
\end{figure}

When measurement noise exists in $\overline{\Delta}(s)$, some misidentified cases appear. However, our identification method still works well. For example, when we add $100\%$ noise to $\overline{\Delta}(s)$, only about $13\%$ of total $140$ cases are misidentified on average. In the following, we list four observations.
\begin{enumerate}
    \item When more noise is added to $\overline{\Delta}(s)$, more misidentified cases happen. We can observe such trend from Fig.~\ref{fig:numMidIDVsNoise}.
    \item For the same level of noise, a damage case which happens at a deeper
    generation is more likely to be misidentified.
    \item For the same level of noise, and for the damage cases which happen at the same generation, those cases occurring at inner components are more inclined to misidentification compared to those occurring at outer components.
    \item For the same level of noise, and for the same damaged component, misidentification happens more frequently when the damage amount $\epsilon$ is close to $1$ (undamaged).
\end{enumerate}

\begin{figure}
    \centering
    \includegraphics[width=0.48\textwidth]{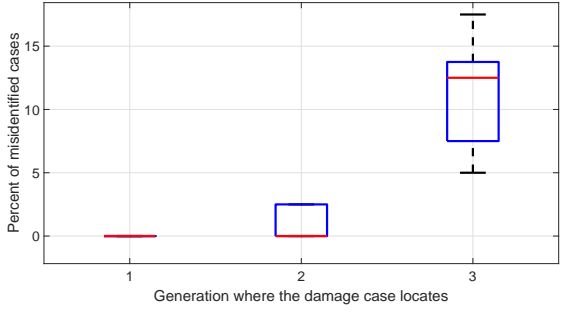}
    \caption{Box plot for the percentage of the misidentified cases at each generation. Ten runs when $50\%$ noise is added.}
    \label{fig:misIDVsGen}
\end{figure}

\begin{figure}
    \centering
    \includegraphics[width=0.48\textwidth]{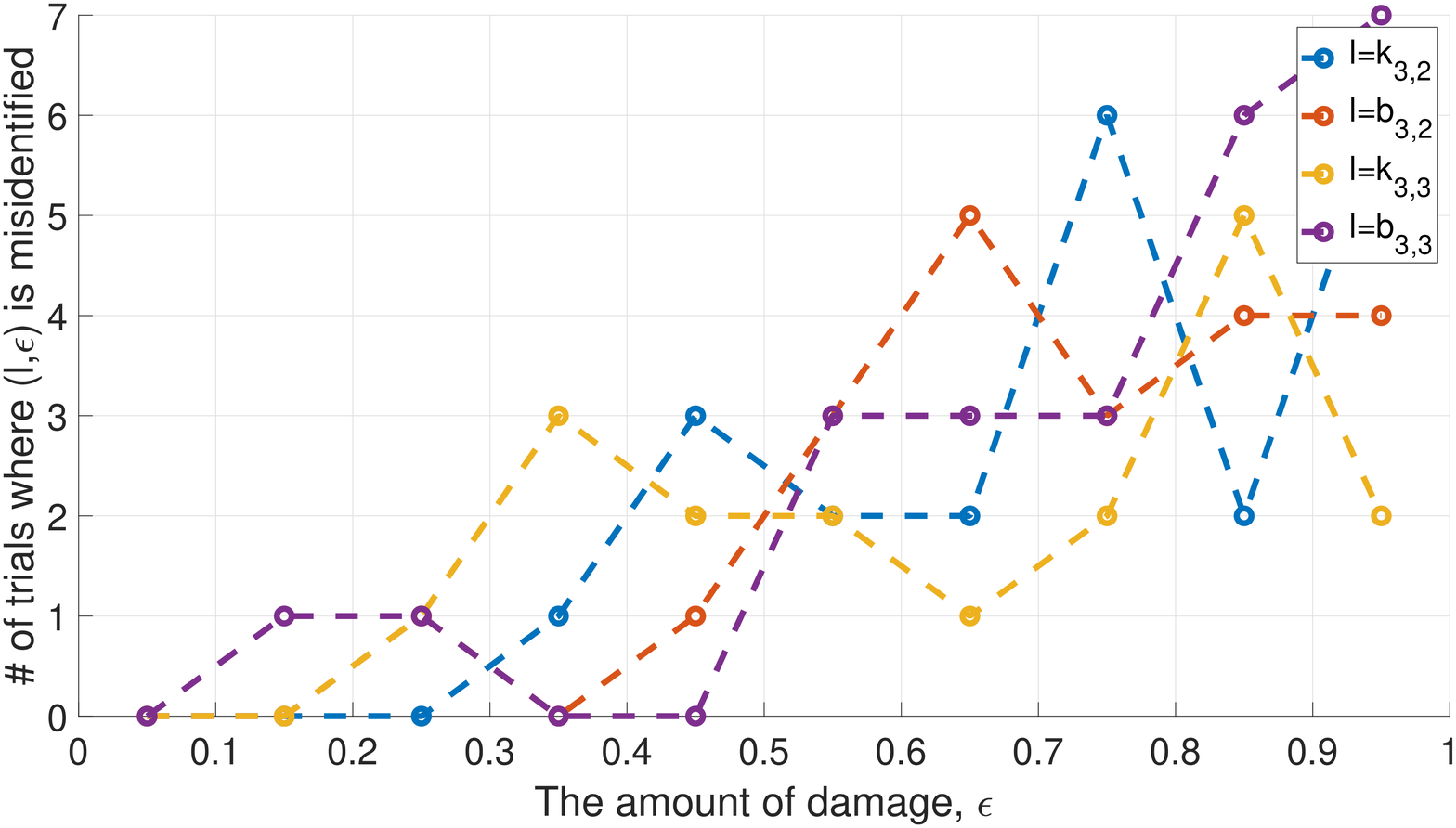}
    \caption{Number of trials out of 10 total trials where each damage case ($l$, $\epsilon$) is misidentified \textit{versus} the amount of damage $\epsilon$.}
    \label{fig:misIDVsE}
\end{figure}

The above observations from (2) to (4) are shown next based on ten different runs where $50\%$ noise is added to $\overline{\Delta}(s)$. Fig.~\ref{fig:misIDVsGen} shows the percentage of the misidentified cases at each generation, from which we can confirm the observation~(2).
For the observation~(3), we focus on the third generation where we call $k_{3,1}$, $b_{3,1}$, $k_{3,4}$, $b_{3,4}$ outer components, and $k_{3,2}$, $b_{3,2}$, $k_{3,3}$, $b_{3,3}$ inner components. Table~\ref{tab:innerOuter} shows average percentage of misidentified cases for components on the third generation, from which we can confirm the observation~(3).
\begin{table}
    \centering
    \caption{Average percentage of misidentified cases for components on the third generation.}
    \begin{tabular}{|c|c|}
        \hline
        Damaged components & Average $\%$ of misidentified cases \\ \hline
        $k_{3,1}$, $b_{3,1}$ & $0.5\%$ \\ \hline
        $k_{3,2}$, $b_{3,2}$, $k_{3,3}$, $b_{3,3}$ & $20.75\%$ \\ \hline
        $k_{3,4}$, $b_{3,4}$ & $2\%$ \\ \hline
    \end{tabular}
    \label{tab:innerOuter}
\end{table}
For the observation~(4), we focus on those inner components at the third generation which have most misidentified cases. Fig.~\ref{fig:misIDVsE} plots the number of trials out of total 10 runs where each damage case ($l$, $\epsilon$) is misidentified \textit{versus} the amount of damage $\epsilon$, from which we can confirm the observation~(4).

\begin{figure}
    \centering
    \includegraphics[width=0.48\textwidth]{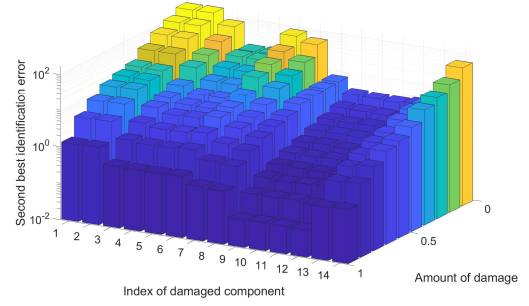}
    \caption{Second best identification error \textit{versus} the index of damaged component $l$ and the amount of damage $\epsilon$.}
    \label{fig:secondBest_3D}
\end{figure}

Finally, to summarize the observations from (2) to (4), we plot the second best identification error against both the index of the damaged component $l$ and the amount of damage $\epsilon$ in Fig.~\ref{fig:secondBest_3D}. Note that Fig.~\ref{fig:secondBest_3D} is based on only one of those ten trials where $50\%$ noise is added to $\overline{\Delta}(s)$, but plots for all the other nine trials are qualitatively similar. Recall that in our identification procedure, we solve an optimization problem~(\ref{eq:optimization}) to find the minimum identification error at each component. Then, among all those locally best identification errors, the globally smallest one gives the final identification result. Therefore, for a damage case ($l$, $\epsilon$), how small its globally second best identification error is indicates how easily that damage case can be misidentified. Here, in order to make plotting convenient, we index the components up to the third generation by integers from 1 to 14, that is, $k_{1,1}\rightarrow1$, $b_{1,1}\rightarrow2$, $\dots$, $b_{3,4}\rightarrow14$. Hence, indices $\{1,2\}$ represent components on the first generation, indices $\{3,4,5,6\}$ represent components on the second generation, and indices $\{7,8,\dots,14\}$ represent components on the third generation. Moreover, $\{9,10,11,12\}$ represent the inner components on the third generation. Therefore, from Fig.~\ref{fig:secondBest_3D}, we can get an overall idea for the observations (2) to (4).

\section{Discussion}
\label{sec:discussion}

\subsection{Justification for using piecewise linear interpolation to make the mapping from $\epsilon$ to $z_j$ and $p_j$ continuous}

As described in Section~\ref{sec:id}, before the damage identification, we sample $z_j$ and $p_j$ at discrete $\epsilon_a$ which is picked as stated by Eq.~(\ref{eq:cheb}). Then, that mapping is made continuous using piecewise linear interpolation so that $z_j$ and $p_j$ can be computed for all $0<\epsilon<1$ according to Eqs.~(\ref{eq:zContinuous}) and~(\ref{eq:pContinuous}).

\begin{figure}
    \centering
    \includegraphics[width=0.48\textwidth]{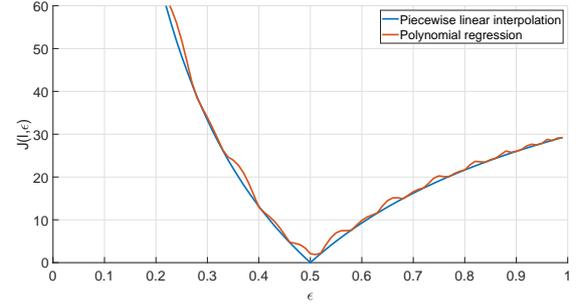}
    \caption{Comparison between two $J(k_{2,1},\epsilon)$ where the measured $\overline{\Delta}(s)$ is for $(l,\epsilon)=(k_{2,1},0.5)$. The blue one is obtained by using piecewise linear interpolation to construct a continuous mapping from $\epsilon$ to $z_j$ and $p_j$ based on those sampled values $z_j(k_{2,1},\epsilon_a)$ and $p_j(k_{2,1},\epsilon_a)$. The red one uses the polynomial regression to construct that continuous mapping.}
    \label{fig:polyFit_JCompare}
\end{figure}

The most important consideration of picking a suitable interpolation method in this case is that it does not create extra saddle points and local minima in the identification error $J(l,\epsilon)$, since those would prevent the decision variable from converging to the actual minimizer. Fig.~\ref{fig:polyFit_JCompare} compares two different $J(k_{2,1},\epsilon)$ \textit{versus} $\epsilon$ where the $\overline{\Delta}(s)$ is for the damage case $(l,\epsilon)=(k_{2,1},0.5)$. One $J(k_{2,1},\epsilon)$ is obtained by using piecewise linear interpolation to construct a continuous mapping from $\epsilon$ to $z_j$ and $p_j$ based on those sampled values $z_j(k_{2,1},\epsilon_a)$ and $p_j(k_{2,1},\epsilon_a)$. The other $J(k_{2,1},\epsilon)$ is obtained by using polynomial regression to construct that continuous mapping. From Fig.~\ref{fig:polyFit_JCompare}, we see that the $J(k_{2,1},\epsilon)$ obtained by polynomial regression has a lot of oscillations. Such behavior is due to the oscillatory nature of polynomial regression known as Runge's phenomenon, and it is undesirable. Should another interpolation method lead to a smooth $J(l,\epsilon)$ similar to the blue curve in Fig.~\ref{fig:polyFit_JCompare}, it can also be used.

\subsection{Effects brought by a deeper damage}
\begin{figure}
    \centering
    \includegraphics[width=.48\textwidth]{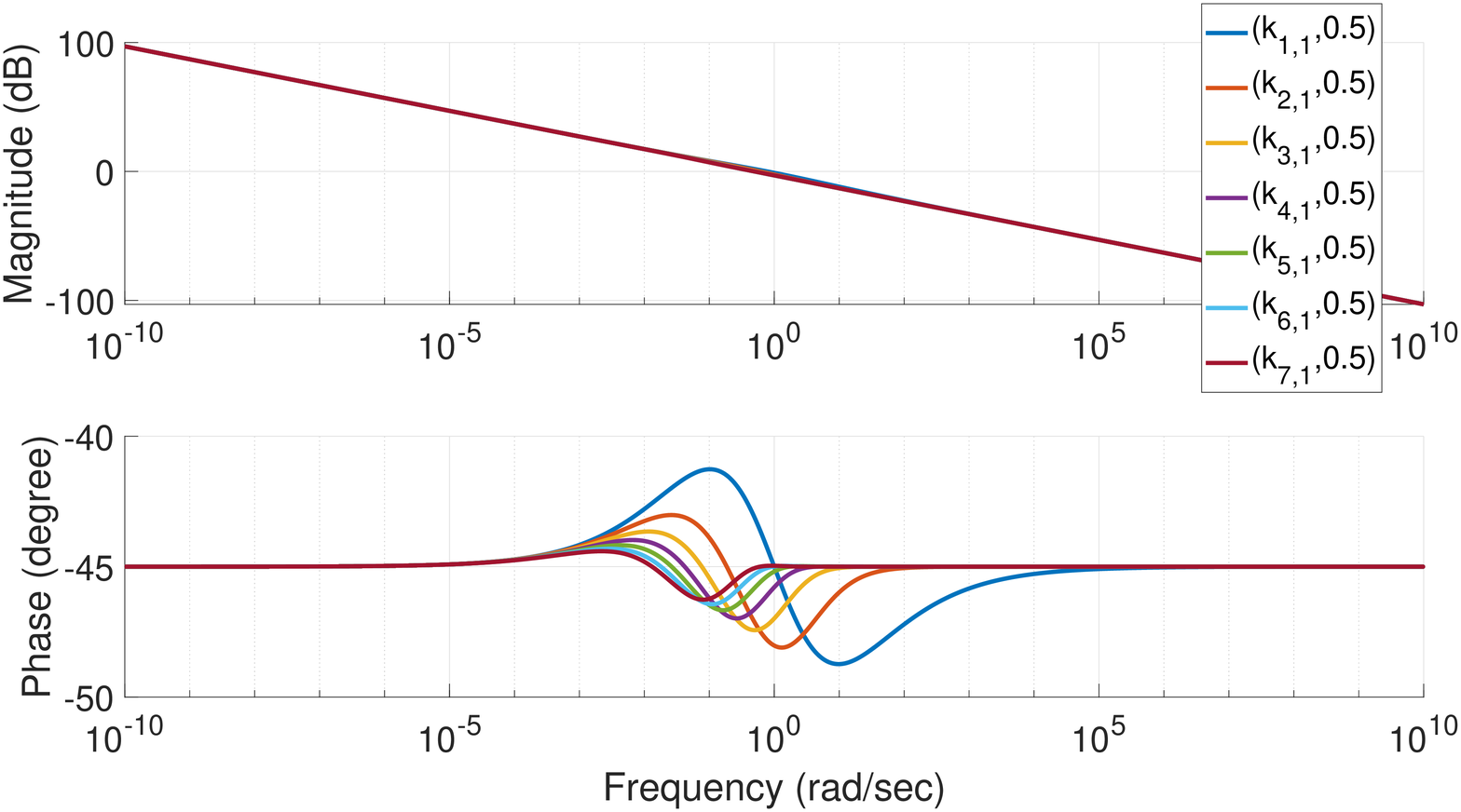}
    \caption{The tree's overall frequency responses for $7$ different damage cases where the damaged component is $k_{1,1}$ through $k_{7,1}$, and the damage amount stays at $0.5$.}
    \label{fig:deeper}
\end{figure}

There are three aspects of consequences when a damage goes deeper inside the tree. First, a deep damaged component is naturally difficult to be identified as it has little effect on the overall frequency response of the tree. Fig.~\ref{fig:deeper} plots the frequency response of the same damage amount when that damage goes from the first generation to the seventh, from which we can see that the discrepancy between two curves is less obvious for a deeper damage.

Second, a deeper damage requires a larger database to store discrete samples of $z_j$ and $p_j$, which thus also requires more time to construct that database. Note that from the first to the $n$-th generation, there are $2^{n+1}-2$ components in total. Therefore, the size of that database doubles for each generation deeper. As for the time required to construct that database, we note that the computation of a damage case is always based on other damage cases which have already been stored in that database. Therefore, the computation time of a new damage case would not be significantly affected by its generation. As a result, the total time consumption to construct that database is proportional to its size, and thus it would also double for each generation deeper.

Third, a deeper damage requires more iterations during our identification procedure, because the size of $L$ increases. As a result, the outer \texttt{for} loop in Algorithm~\ref{alg:1} iterates more. For the same reason explained in the above paragraph, the size of $L$ doubles for each generation deeper, so the total running time would also double. Note that the time consumption inside the outer \texttt{for} loop is independent of how deep a damage case is.

\section{Conclusion}
\label{sec:conclusion}

In this paper, we propose a method to identify the damaged component $l$ and quantify its damage amount $\epsilon$ in a damaged tree given its overall frequency response. Our identification procedure iterates through all possible components and solve a nonlinear programming problem~(\ref{eq:optimization}) at each component. Formulation of that optimization problem takes advantage of our previous work about modeling the damaged tree model as $G_\infty(s)\cdot\Delta_{(l,\epsilon)}(s)$ where $\Delta_{(l,\epsilon)}(s)$ is completely determined by its corresponding damage case ($l$,$\epsilon$) as shown in Eq.~(\ref{eq:delta}). In addition, the performance of that identification algorithm and the effects brought by a damaged component at a deep generation are also discussed in this paper.


\bibliography{ifacconf}             

\end{document}